\documentclass[11pt,twoside]{article}
\usepackage{./asp2014}
\aspSuppressVolSlug
\resetcounters

\begin{document}
\title{Instabilities in Interacting Binary Stars}
\author{I.~L.~Andronov$^{1}$, 
        K.~D.~Andrych$^{2}$,
        K.~A.~Antoniuk$^{3}$,
        A.~V.~Baklanov$^{3}$,
        P.~Beringer$^{4}$,
        V.~V.~Breus$^{1}$,
        V.~Burwitz$^{5}$,
        L.~L.~Chinarova$^{2}$, 
        D.~Chochol$^{6}$, 
        L.~M.~Cook$^{7}$,
        M.~Cook$^{8}$,
        P.~Dubovsk\'y$^{9}$, 
        W.~God{\l}owski$^{10}$,
        T.~Heged\"us$^{4}$,
        K.~Ho\v{n}kov\'a$^{11}$,
        L.~Hric$^{6}$, 
        Young-Beom Jeon$^{12},$
        J.~Jury\v{s}ek$^{11,13}$,
        Chun-Hwey Kim$^{14}$,
        Yonggi Kim$^{14}$,
        Young-Hee Kim$^{14}$,
        S.~V.~Kolesnikov$^{2,3}$,
        L.~S.~Kudashkina$^{1}$,
        A.~V.~Kusakin$^{15}$,
        V.~I.~Marsakova$^{2}$,
        P.~A.~Mason$^{16}$, 
        M.~Ma\v{s}ek$^{11,13}$,
        N.~Mishevskiy$^{17}$,
        R.~H.~Nelson$^{18}$,
        A.~Oksanen$^{19}$,
        S.~Parimucha$^{20}$,
        Ji-Won Park$^{14}$,
        K.~Petr\'ik$^{21}$,
        C.~Qui\~{n}ones$^{22}$,
        K.~Reinsch$^{23}$,
        J.~W.~Robertson$^{24}$,
        I.~M.~Sergey$^{25}$, 
        M.~Szpanko$^{10}$,
        M.~G.~Tkachenko$^{1}$,
        L.~G.~Tkachuk$^{26}$, 
        I.~Traulsen$^{27}$,
        J.~Tremko$^{6}$, 
        V.~S.~Tsehmeystrenko$^{28}$,
        Joh-Na Yoon$^{14}$,
        S.~Zo{\l}a$^{29,30}$,\\
\fbox{N.~M.~Shakhovskoy}$^{3}$
\affil{$^1$ Department of Mathematics, Physics and Astronomy, Odessa National Maritime University, Odessa, Ukraine; \email{tt\_ari@ukr.net}}
\affil{$^2$ Department of Theoretical Physics and Astronomy; Astronomical Observatory, I.I. Mechnikov Odessa National University, Odessa, Ukraine}
\affil{$^3$ Crimean Astrophysical Observatory, Nauchny, Crimea}
\affil{$^4$ Baja Astronomical Observatory of Szeged University, Baja,Hungary}
\affil{$^5$ Max-Planck-Institut f\"ur extraterrestrische Physik, M\"unchen, Germany}
\affil{$^6$ Astronomical Institute of the Slovak Academy of Sciences, Tatranska Lomnica, Slovakia}
\affil{$^7$ CBA Concord, 1730 Helix Ct., Concord CA 94518}
\affil{$^8$ AAVSO, Newcastle Observatory, Newcastle, Ontario, Canada}
\affil{$^9$ Vihorlat Observatory Humenn\'e, Slovakia}
\affil{$^{10}$ Institute of Physics, Opole University, Opole, Poland}
\affil{$^{11}$ Variable Star and Exoplanet Section of Czech Astronomical Society}
\affil{$^{12}$ Korea Astronomy Observatory and Space Science Institute, Daejeon, Korea}
\affil{$^{13}$ Institute of Physics of the Czech Academy of Sciences, 
Praha, Czechia}
\affil{$^{14}$ Chungbuk National University, Cheongju, Korea}
\affil{$^{15}$ V.G. Fesenkov Astrophysical institute, Almaty, Kazahstan}
\affil{$^{16}$ New Mexico State University, Las Cruces, NM, USA}
\affil{$^{17}$ Private observatory Ananjev (L33)}
\affil{$^{18}$ Guest investigator, Dominion Astrophysical Observatory}
}
\pagebreak

\author{
\affil{$^{19}$ Hankasalmi Observatory, Jyv\"askyl\"a, Finland}
\affil{$^{20}$ Institute of Physics, Faculty of Science, University of P.J. \v{S}af\'{a}rik, Ko\v{s}ice, Slovakia}
\affil{$^{21}$ Astronomical Observatory and Planetarium, Hlohovec, Slovakia}
\affil{$^{22}$ Universidad Nacional de C\'{o}rdoba, IATE, C\'{o}rdoba, Argentina}
\affil{$^{23}$ Georg--August University, G\"ottingen, Germany}
\affil{$^{24}$ Department of Physical Sciences, Arkansas Tech University, Russellville}
\affil{$^{25}$ Amateur society "Astroblocknote", Minsk, Belarussia}
\affil{$^{26}$ Private Observatory, Kiev, Ukraine}
\affil{$^{27}$ Leibniz-Institut f\"ur Astrophysik Potsdam (AIP), Potsdam, Germany}
\affil{$^{28}$ Private Observatory "Heavenly Owl", Odessa, Ukraine}
\affil{$^{29}$ Jagiellonian University, Krak\'ow, Poland}
\affil{$^{30}$ Pedagogical University, Krak\'ow, Poland}
}

\paperauthor{I.L.~Andronov}{tt\_ari@ukr.net}{orcid.org/0000-0001-5874-0632}{Odessa National Maritime University}{Department of Mathematics, Physics and Astronomy}{Odessa}{Odessa}{65029}{Ukraine}
\paperauthor{K.A.~Andrych}{katyaandrich@gmail.com}{orcid.org/0000-0002-3780-4112}{I.I. Mechnikov Odessa National University}{Department of Theoretical Physics and Astronomy}{Odessa}{Odessa}{65014}{Ukraine}
\paperauthor{K.A.~Antoniuk}{e-mail}{orcid.org/}{Crimean Astrophysical Observatory}{Department}{City}{State/Province}{Postal Code}{Crimea}
\paperauthor{A.V.~Baklanov}{alex-baklanov@mail.ru}{orcid.org/0000-0001-7684-059X}{Crimean Astrophysical Observatory}{Department}{City}{State/Province}{Postal Code}{Crimea}
\paperauthor{P.~Beringer}{e-mail}{orcid.org/}{Baja Astronomical Observatory of Szeged University}{Department}{Baja}{State/Province}{Postal Code}{Hungary}
\paperauthor{V.V.~Breus}{bvv_2004@ua.fm}{orcid.org/0000-0002-7535-2241}{Odessa National Maritime University}{Department of Mathematics, Physics and Astronomy}{Odessa}{Odessa}{65029}{Ukraine}
\paperauthor{V.~Burwitz}{burwitz@mpe.mpg.de}{orcid.org/}{Max-Planck-Institut f\"ur extraterrestrische Physik}{Department}{M\"unchen}{State/Province}{85741}{Germany}
\paperauthor{L.L.~Chinarova}{lidia_chinarova@mail.ua}{http://orcid.org/0000-0002-4684-7122}{I.I. Mechnikov Odessa National University}{Astronomical Observatory}{Odessa}{Odessa}{65014}{Ukraine}
\paperauthor{D.~Chochol}{chochol@ta3.sk}{orcid.org/}{Astronomical Institute of the Slovak Academy of Sciences}{Department}{Tatranska Lomnica}{State/Province}{Postal Code}{Slovakia}
\paperauthor{L.~M.~Cook}{lew.cook@gmail.com}{orcid.org/0000-0003-3332-9649}{Center for Backyard Astrophysics}{Concord}{Concord}{CA}{94518}{USA}
\paperauthor{M.~Cook}{michael.cook@newcastleobservatory.ca}{orcid.org/0000-0003-1913-7309}{AAVSO, Newcastle Observatory}{Department}{Newcastle}{Ontario}{Postal Code}{Canada}
\paperauthor{P.~Dubovsk\'y}{var@kozmos.sk}{orcid.org/}{Vihorlat Observatory Humenn\'e}{Department}{Humenn\'e}{State/Province}{Postal Code}{Slovakia}
\paperauthor{W.~God{\l}owski}{godlowski@uni.opole.pl}{orcid.org/}{Opole University}{Institute of Physics}{Opole}{State/Province}{Postal Code}{Poland}
\paperauthor{T.~Heged\"us}{hege@bajaobs.hu}{orcid.org/}{Baja Astronomical Observatory of Szeged University}{Department}{Baja}{State/Province}{Postal Code}{Hungary}
\paperauthor{K. ~Ho\v{n}kov\'a}{katerina.honkova@astronomie.cz}{orcid.org/}{Variable Star and Exoplanet Section of Czech Astronomical Society}{Department}{City}{State/Province}{Postal Code}{Czech Republic}
\paperauthor{L.~Hric}{hric@ta3.sk}{orcid.org/}{Astronomical Institute of the Slovak Academy of Sciences}{Department}{Tatranska Lomnica}{State/Province}{Postal Code}{Slovakia}
\paperauthor{Young-Beom Jeon}{ybjeon@kasi.re.kr}{orcid.org/}{Korea Astronomy Observatory and Space Science Institute}{Bohyunsan Optical Astronomy Observatory}{Daejeon}{State/Province}{34055}{Korea}
\paperauthor{J.~Jury\v{s}ek}{jakubjurysek@astronomie.cz}{orcid.org/}{Silesian University in Opava}{Institute of Physics, Faculty of Philosophy \& Science}{Opava}{State/Province}{Postal Code}{Czech Republic}
\paperauthor{Chun-Hwey Kim}{kimch@chungbuk.ac.kr>}{orcid.org/}{Chungbuk National University}{Department}{Cheongju}{State/Province}{Postal Code}{Korea}
\paperauthor{Yonggi Kim}{ykkim153@chungbuk.ac.kr}{orcid.org/}{Chungbuk National University}{Department}{Cheongju}{State/Province}{Postal Code}{Korea}
\paperauthor{Young-Hee Kim}{younghekim24@gmail.com}{orcid.org/}{Chungbuk National University}{Department}{Cheongju}{State/Province}{Postal Code}{Korea}
\paperauthor{S.V.~Kolesnikov}{s_v-k@mail.ru}{orcid.org/}{I.I. Mechnikov Odessa National University}{Astronomical Observatory}{Odessa}{Odessa}{65014}{Ukraine}
\paperauthor{S.V.~Kolesnikov}{s_v-k@mail.ru}{orcid.org/}{Crimean Astrophysical Observatory}{Author2 Department}{City}{Nauchny}{Postal Code}{Crimea}
\paperauthor{L.S.~Kudashkina}{kudals04@mail.ru}{orcid.org/0000-0002-8482-9240}{Odessa National Maritime University}{Department of Mathematics, Physics and Astronomy}{Odessa}{Odessa}{65029}{Ukraine}
\paperauthor{A.V.~Kusakin}{un7gbd@gmail.com}{orcid.org/0000-0002-7756-546X}{V.G. Fesenkov Astrophysical institute}{}{Almaty}{}{050020}{Kazahstan}
\paperauthor{V.I.~Marsakova}{v.marsakova@onu.edu.ua}{orcid.org/0000-0001-6306-1781}{I.I. Mechnikov Odessa National University}{Department of Theoretical Physics and Astronomy}{Odessa}{Odessa}{65014}{Ukraine}
\paperauthor{P.A.~Mason}{pmason@nmsu.edu}{orcid.org/}{New Mexico State University}{Department}{Las Cruces}{New Mexico }{Postal Code}{USA}
\paperauthor{M.~Ma\v{s}ek}{cassi@astronomie.cz}{orcid.org/}{Variable Star and Exoplanet Section of Czech Astronomical Society}{Department}{City}{State/Province}{Postal Code}{Czech Republic}
\paperauthor{M.~Ma\v{s}ek}{e-mail}{orcid.org/}{Institute of Physics of the Czech Academy of Sciences}{Department}{Praha}{State/Province}{Postal Code}{Czech Republic}
\paperauthor{N.~Mishevskiy}{nikastro@ukr.net}{orcid.org/0000-0001-8860-5861}{Private observatory Ananjev (L33)}{}{Ananjev}{Odessa}{66400}{Ukraine}
\paperauthor{R.H.~Nelson}{bob.nelson@shaw.ca}{orcid.org/}{Dominion Astrophysical Observatory}{Department}{City}{State/Province}{Postal Code}{Country}
\paperauthor{A.~Oksanen}{arto.oksanen@jklsirius.fi}{orcid.org/}{Hankasalmi Observatory}{Department}{Jyv\"askyl\"a}{State/Province}{Postal Code}{Finland}
\paperauthor{S.~Parimucha}{parimucha@ta3.sk}{orcid.org/}{University of P.J. \v{S}af\'{a}rik}{Institute of Physics, Faculty of Science}{Ko\v{s}ice}{State/Province}{Postal Code}{Slovakia}
\paperauthor{Ji-Won Park}{jiwon716@gmail.com}{orcid.org/}{Chungbuk National University}{Department}{Cheongju}{State/Province}{Postal Code}{Korea}
\paperauthor{K.~Petr\'ik}{kpetrik@astronyx.sk}{orcid.org/}{Astronomical Observatory and Planetarium}{Department}{Hlohovec}{State/Province}{Postal Code}{Slovakia}
\paperauthor{C.~Qui\~{n}ones}{mececilq@gmail.com}{orcid.org/0000-0003-0276-9879}{Universidad Nacional de C\'{o}rdoba}{IATE}{C\'{o}rdoba}{State/Province}{Postal Code}{Argentina}
\paperauthor{K.~Reinsch}{reinsch@astro.physik.uni-goettingen.de}{orcid.org/}{Georg--August University}{Department}{G\"ottingen}{State/Province}{Postal Code}{Germany}
\paperauthor{J.W.~Robertson}{jrobertson@atu.edu}{orcid.org/}{Arkansas Tech University}{Department of Physical Sciences}{Russellville}{Arkansas}{Postal Code}{USA}
\paperauthor{I.M.~Sergey}{seriv@rambler.ru}{orcid.org/}{Amateur society "Astroblocknote"}{Department}{Minsk}{State/Province}{Postal Code}{Belarussia}
\paperauthor{B.M.~Shergelashvili}{bidzina.shergelashvili@oeaw.ac.at}{orcid.org/}{Space Research Institute}{Department}{Graz}{State/Province}{Postal Code}{Austria}
\paperauthor{M.~Szpanko}{szpanko@uni.opole.pl}{orcid.org/}{Opole University}{Institute of Physics}{Opole}{State/Province}{Postal Code}{Poland}
\paperauthor{M.G.~Tkachenko}{masha.vodn@yandex.ua}{orcid.org/0000-0002-3166-2627}{Odessa National Maritime University}{Department of Mathematics, Physics and Astronomy}{Odessa}{Odessa}{65029}{Ukraine}
\paperauthor{L.G.~Tkachuk}{tlgleonid@bank.gov.ua}{orcid.org/}{Private Observatory}{Department}{Kiev}{State/Province}{Postal Code}{Ukraine}
\paperauthor{I.~Traulsen}{itraulsen@aip.de}{orcid.org/}{Leibniz-Institut f\"ur Astrophysik Potsdam (AIP)}{X-ray Astronomy}{Potsdam}{State/Province}{Postal Code}{Germany}
\paperauthor{J.~Tremko}{tremko@ta3.sk}{orcid.org/}{Astronomical Institute of the Slovak Academy of Sciences}{Department}{Stara Lesna}{State/Province}{Postal Code}{Slovakia}
\paperauthor{V.S.~Tsehmeystrenko}{tseh_valery@ukr.net}{orcid.org/}{Private Observatory "Heavenly Owl"}{Department}{Odessa}{State/Province}{Postal Code}{Ukraine}
\paperauthor{Joh-Na Yoon}{antalece@chungbuk.ac.kr}{orcid.org/}{Chungbuk National University}{Department}{Cheongju}{State/Province}{Postal Code}{Korea}
\paperauthor{S.~Zo{\l}a}{szola@oa.uj.edu.pl}{orcid.org/}{Jagiellonian University}{Department}{Krak\'ow}{State/Province}{Postal Code}{Poland}
\paperauthor{S.~Zo{\l}a}{szola@oa.uj.edu.pl}{orcid.org/}{Pedagogical University}{Department}{Krak\'ow}{State/Province}{Postal Code}{Poland}
\paperauthor{N.M.~Shakhovskoy}{e-mail}{orcid.org/}{Crimean Astrophysical Observatory}{Department}{City}{State/Province}{Postal Code}{Crimea}


\begin{abstract}
The types of instability in the interacting binary stars are briefly reviewed. The project "Inter-Longitude Astronomy" is a series of smaller projects on concrete stars or groups of stars. It has no special funds, and is supported from resources and grants of participating organizations, when informal working groups are created. This "ILA" project is in some kind similar and complementary to other projects like WET, CBA, UkrVO, VSOLJ, BRNO, MEDUZA, AstroStatistics, where many of us collaborate. Totally we studied 1900+ variable stars of different types, including newly discovered variables. The characteristic timescale is from seconds to decades and (extrapolating) even more. The monitoring of the first star of our sample AM Her was initiated by Prof. V.P. Tsesevich (1907-1983). Since more than 358 ADS papers were published. In this short review, we present some highlights of our photometric and photo-polarimetric monitoring and mathematical modeling of interacting binary stars of different types: classical (AM Her, QQ Vul, V808 Aur = CSS 081231:071126+440405, FL Cet), asynchronous (BY Cam, V1432 Aql), intermediate (V405 Aql, BG CMi, MU Cam, V1343 Her, FO Aqr, AO Psc, RXJ 2123, 2133, 0636, 0704) polars and magnetic dwarf novae (DO Dra) with 25 timescales corresponding to different physical mechanisms and their combinations (part "Polar"); negative and positive superhumpers in nova-like (TT Ari, MV Lyr, V603 Aql, V795 Her) and many dwarf novae stars ("Superhumper"); eclipsing "non-magnetic" cataclysmic variables(BH Lyn, DW UMa, EM Cyg; PX And); symbiotic systems ("Symbiosis"); super-soft sources (SSS, QR And); spotted (and not spotted) eclipsing variables with (and without) evidence for a current mass transfer ("Eclipser") with a special emphasis on systems with a direct impact of the stream into the gainer star's atmosphere, which we propose to call "Impactor" (short from "Extreme Direct Impactor"), or V361 Lyr-type stars. Other parts of the ILA project are "Stellar Bell" (interesting pulsating variables of different types and periods - M, SR, RV Tau, RR Lyr, Delta Sct with changes of characteristics) and "Novice"(="New Variable") discoveries and classification based on special own observations and data mining with a subsequent monitoring for searching and studying possible multiple components of variability. Special mathematical methods have been developed to create a set of complementary software for statistically optimal modeling of variable stars of different types.
\end{abstract}

\hfill \parbox{12cm}{
\em "Why carry out the monitoring? Imagine that You made a photo of a falling stone. Only one photo. Then somebody looks at this photo and thinks: "Wow! The stone is flying? Anti-gravitation?  A thin wire?" So, to make physically correct models, You should know the dynamics of the object! So, take a small telescope (nobody will allow You to use a large one for a long time) and observe! As long as possible! And You will catch Your Discovery!\\
$~ $\hfill Prof. Vladimir P. Tsesevich (1907-1983) }

\section{Introduction}\label{intro}

In the epigraph, we see an unofficial justification of a necessity of long-term monitoring of stars. Even so-called "periodic" stars, which are typically believed to exhibit stable light curves, show deviations which may be interpreted by period variations or a light - time effect (cf. \cite{tses71}, \cite{kreiner}). For cataclysmic variables, there are many physical processes with characteristic times from 0.1s to (theoretically) billions of years (cf. \cite{warner}, \cite{a07a}). 
Thus a long -- term monitoring may lead not to a better statistical accuracy of the light curve, but also to a detection of rare or "very rare" events in these systems. The monitoring of the first star of our sample -- AM Her -- was initiated by Prof. V.P.Tsessevich (1907-1983). Since more than 358 papers (cited in the ADS) were published, with a total number of studied variable stars exceeding 1900.

\section{``Inter-Longitude Astronomy'' (ILA)}\label{ila}

The project "Inter-Longitude Astronomy" is a joint name of a series of smaller projects on observation and interpretation of concrete variable stars or groups of stars. It has no special funds, professional and amateur astronomers take part based on their own resources, when informal astronomical working groups are created. This "ILA" project is in some kind similar and complementary to other projects like WET, CBA, UkrVO, VSOLJ, BRNO, MEDUZA and currently Astroinformatics (\cite{vav2016}), where many of us take part. Previous reviews on the ILA project were published by \cite{a03} and \cite{a10}.
The main parts of this project are briefly described below.

\subsection{``Polar''}
Photo-polarimetric and spectroscopic study of gravi-magnetic rotators in cataclysmic variables -  classical, asynchronous and intermediate polars;

The photometric monitoring of AM Her started in Odessa based on photographic observations and later continued photopolarimetrically at the 2.6m and 1.25m telescopes of the Crimean astrophysical observatory (\cite{ES82}, \cite{sak94}). 

The characteristics of the brightness variations of AM Her are studied at time scales from seconds to decades by using the scalegram method proposed by \cite{a97}. The unbiased scatter estimate increases with the filter half-width according to a power law with an index 0.180 from $10^{-4}$ to 3000 days indicating a fractal nature of irregular variations with a low dimension of $D=0.32$ (\cite{aks97}). \cite{a87b} discussed different types of instability of the accretion column.
\cite{ASK03} reviewed methods and results of the principal components of variability of the accretion structures near white dwarfs. \cite{s99} reported on a correlation between the "shot noise" decay time and circular polarization. 
\cite{s96} estimated the temperature of the white dwarf of $\sim20\,000$K with an additional component of $\sim35\,000$K in the bright state, and a magnetic field strength of $\sim13$ MG.
\cite{a03} discovered two-component "shot noise" in the X-Ray variability with decay times 170s and 10s, indicating a MHD instability in the accretion column.
An unprecedented UV Cet-type flare was detected in AM Her (\cite{saak93}, \cite{sak94}). 
More frequent flares with a red spectral energy distribution are due to accretion events (\cite{bb00}).  

Theoretical models of magnetic cataclysmic b³nary systems include: determination of the orientation of the magnetic axis of a white dwarf from orbital curves of the X-Ray flux (\cite{a86}), oscillations of the orientation of the white dwarf ("swingings", "librations") (\cite{a87a}, \cite{ak96},
\cite{Ma02}), "asymmetric propeller" (\cite{a87c}).

\subsection{Eclipsing and Asynchronous Polars}
\cite{aa14} estimated the size of the emitting region in eclipsing polar V808 Aur of $\sim1300$km, which is much smaller than the white dwarf.

\cite{ab07} determined the capture radius and synchronization time of the white dwarf in the unique magnetic cataclysmic system V1432 Aql
-- the asynchronous polar with a spin-up.
\cite{a08} reported an analysis of the idling magnetic white dwarf in the synchronizing polar BY Cam based on the "Noah-2" project. \cite{Ma2013} discussed high speed optical observations of cataclysmic variables: FL Ceti, BY Cam, and DQ Her.

\subsubsection{Intermediate Polars}
For EX Hya, the spin-up characteristic time of $\tau=4.7$ million years was determined (\cite{a2013a}). The list of moments of spin maxima was compiled by \cite{br2013c}. 
 Much faster spin-up of $\tau=170$ thousand years was detected in MU Cam, the modulation of the spin phases with the orbital phase was discovered, being interpreted by periodic changes of the accretion structure due to rotation of the magnetic white dwarf (\cite{p2015}).
 During long-term monitoring, \cite{a2014a} have detected a return to high luminosity state of V1323 Her = RXS J180340.0+401214 in 2014. Similar phenomenon was monitored in the "King of Intermediate polars" -- FO Aqr (in preparation), which shows alternate spin period variations (\cite{br2012}). A complicated period variation was found in BG CMi, where a fourth-order polynomial is the statistically best fit for the timeing of the spin maxima, indicating a third-order polynomial in the period variations, and thus a decreasing rate of the spin-up of the white dwarf (\cite{k05a}). Precession of the white dwarf or a light-time effect due to the third body are responsible on the $O-C$ variations in V405 Aur (\cite{br2013a}).

\cite{h2014} have corrected the orbital and spin periods of a low-field intermediate polar V709 Cas.
\cite{br2015} corrected the orbital period and found evidence for 2-day periodicity of the intermediate polar V2306 Cygni.
\cite{a08b} discovered a correlation between the brightness at the outburst with it's slope in the outbursting intermediate polar (= magnetic dwarf Nova) DO Dra, and also "Transient Periodic Variations" (TPO). 

\subsection{``Stellar Bell''}

Analysis of multi-component pulsations of short- and long- period variable stars based on own photometric observations and the data from the international databases of AFOEV (France), VSOLJ (Japan) and AAVSO (USA). Recent review of our results was presented by \cite{a2014b}. \cite{ac2013} reviewed the method of "Running Sines" as a powerful tool to study periodic and aperiodic modulations of the nearly sinusoidal shape of oscillations.
\cite{a2012c} studied two-component variability of the semi-regular pulsating star U Del.
\cite{m2012} discussed more transient type variables -- Miras or SRa'a. The secular variations of the photometric parameters of Mira Ceti variables and semiregular variables were studied by \cite{m2013b}, \cite{m2014a} and \cite{m2014b}.
\cite{ku2016a} analyzed  mean light curves of the Mira-type stars in the H- and K-bands.
\cite{ku2012} discussed the classification of semiregular pulsating asymptotic giants branch stars.

\subsection{"Eclipser" - "Clean" and "Spotted" Eclipsing Variable Stars}

Monitoring and analysis of binaries with spots with a main attention to self-consistent modeling of systems with migrating spots and "direct impactors" with accretion - induced "hot spots". 

The algoritm NAV ("New Algol Variable"), originally introduced for phenomenological modelling of the light curves by \cite{a2012a}, was used 
for determination of characteristics of newly discovered eclipsing binary 
(\cite{a2012b}, \cite{tk2015}), while \cite{tk2016b} determined these parameters for the prototype eclipsing binaries Algol, $\beta$ Lyrae and W UMa. Numerous possible improvements of the method NAV by introducing a minimal number of additional phenomenological parameters were introduced (\cite{a2016a}, \cite{a2016b}, \cite{a2017}).
\cite{a2015a} used phenomenological parameters for the two-color photometry to estimate physical parameters of VSX J180243.9+400331. 

The "simplified" model assuming spherical components was analyzed (\cite{a2013e}, \cite{tk2014}). \cite{tk2016a} analyzed the structure of the corresponding test function.
 
The classical $O-C$ analysis of the minima timings of eclipsing variables was resulted in detection of cyclic period changes in the EB-type systems KR Cyg, V382 Cyg and BX And (\cite{tv2015}).

\subsection{Extreme Direct Impactors (V361 Lyr-type stars)}

An exotic group of eclipsing binary systems currently contains only two objects - V361 Lyr, the model of which was proposed by \cite{ar87}, and of a recently discovered object =VSX J052807.9+725606 (\cite{vaa}), which was almost registered in the GCVS (\cite{gcvs}) as V549 Cam. These objects show an extremely large difference in maxima, which are interpreted by a bright spot caused by a direct impact of the accretion stream in the atmosphere of the secondary without a creation of the accretion disk. These stars are suggested to be at a pre-contact stage and evolve towards an EW-type binaries. The small number of such known systems argues for s short duration of this evolutionary stage.

\subsection{Other Types of Variability}
Among the variable stars studied within the "Inter-Longitude Astronomy" project, there are objects of many types, among them: 
the symbiotic stars (the direction called the "Symbiosis");
"Superhump" - study of the precession of accretion disks in the Nova-Like (NL) and Dwarf Nova (DN) stars.
The main objects of this type in our study were MV Lyr (\cite{a92}) and TT Ari (\cite{k09}, \cite{a99});
few year - scale variations of characteristics of the outburst cycle length in dwarf novae (\cite{as90}) due to modulation of the accretion rate caused by a solar-type activity of the red dwarf and/or a presence of the third body (\cite{ac96});
Super-Soft Sources V Sge and QR And;
new variable stars ("Novice": star classification and justification of suspected and newly discovered variables from surveys and own observations). 
Time series analysis based on own and published photometric observations obtained in ground-based and space observatories.

\subsection{Software for Processing Astronomical Observations of Variable Stars}\label{math}
Special mathematical methods have been developed to create a set of complementary software for statistically optimal modeling of variable stars of different types.

For analysis of the raw photopolarimetric observations obtained using the 2.6m Shain telescope of the Crimean astrophysical observatory, the software "PolarObs" was developed (\cite{br07}, \cite{k2016}), which was used not only for observations of cataclysmic binaries (\cite{a07a}), but also asteroids and comets (\cite{kol2012}).

\cite{ab04} elaborated a program "MCV" (Multi-Column Viewer), which may be used for multiple purposes, starting from a simple visualisation of the multiple columns to advanced modeling of the multi-periodic multi-harmonic signals with a polynomial trend, corresponding periodogram analysis and improving CCD photometry using the algorithm of the "artificial star".

\cite{ay2015} developed a program for determination of the characteristics of extrema using the statistically optimal approximation among a set of symmetric (typically for eclipsing binaries) and asymmetric (for pulsating variables). Some of these approximations were described by \cite{a87d},

\cite{am2013} compared different methods for minimization of the test function. Besides, for the method of Monte-Carlo, the number of iterations for a better result is statistically the same as the number of computations before this "currently best" value was obtained. 

\cite{a2013d} presented a program WWZ for visualisation of results of the wavelet analysis of astronomical signals with irregularly spaced arguments based on the algoritm of \cite{a98wwz}.

\cite{ac2013} reviewed the base and application of the method of "Running Sines", which is related to Morlet-type wavelet analysis improved for irregularly spaced data (\cite{a98wwz}). The method is most effective for "nearly periodic" signals, which exhibit a "wavy shape" with a "cycle length" varying within few dozen per cent.
A new method to investigate periods and their variation is proposed.

\subsection{On-Going Studies and Future Plans ("todo" list)}\label{todo}
As seen from the "highlights", the monitoring of variable stars produces many interesting results, which could not be obtained at larger telescopes because of high competition for the observational time and thus relatively short time for a given object. The "Inter-Longitude Astronomy" project allows, in some cases, to get data from different longitudes and to solve problems of cycle numbering etc. Although, the "next step" results are determined in our studies, the continuation of monitoring will lead to some expected (and hopefully "unexpected"=discoveries) results, namely search for time evolution of periods and other characteristics of variability. Among the key objects, we plan to continue monitoring of:
\begin{itemize}
\item Classical Polars (AM Her, QQ Vul) - polarimetrically at the 2.6m telescope
\item Classical Eclipsing Polars (FL Cet, V808 Aur = OTJ 0711) 
\item Asynchronous Eclipsing Polar (V1432 Aql)
\item Intermediate Polars (FO Aqr, AO Psc, PQ Gem, MU Cam, V1343 Aql, V1309 Ori, V709 Cas, BG CMi,  V647 Aur=RXJ0636, V418 Gem=RXJ0704, DQ Her)
\item Eclipsing stars with the O'Connell effect (V0452 Dra=VSX J112825+683717, WZ Crv, NOMAD-1 1617-0165649=V1015 Cep, BS Cas, V1094 Cas)
\item Extreme Direct Impactors (V361 Lyr, V549 Cam=VSX J052807.9+725606)
\end{itemize}

\end{document}